\documentclass[conference]{IEEEtran}
\usepackage{cite}
\usepackage{amsmath,amssymb,amsfonts}
\usepackage{algorithmic}
\usepackage{graphicx}
\usepackage{textcomp}
\usepackage{xcolor}
\def\BibTeX{{\rm B\kern-.05em{\sc i\kern-.025em b}\kern-.08em
    T\kern-.1667em\lower.7ex\hbox{E}\kern-.125emX}}
    
\usepackage[hyphens]{url}
\usepackage[hidelinks]{hyperref}
\hypersetup{breaklinks=true}
\begin{document}

\title{Simulation of Medium-Frequency R-Mode Signal Strength}

\author{\IEEEauthorblockN{Jaewon Yu} 
\IEEEauthorblockA{\textit{School of Integrated Technology} \\
\textit{Yonsei University}\\
Incheon, Republic of Korea \\
jaewon.yu@yonsei.ac.kr}
\and
\IEEEauthorblockN{Joon Hyo Rhee${}^{*}$}
\IEEEauthorblockA{\textit{Korea Research Institute of Standards and Science} \\
Daejeon, Republic of Korea \\
jh.rhee@kriss.re.kr} 
{\small${}^{*}$ Corresponding author}
}

\maketitle

\begin{abstract}
Assuming failure in the global navigation satellite systems due to radio frequency interference and ionospheric anomaly, an R-Mode system, a terrestrial integrated navigation system, is being actively studied for domestic deployment in South Korea. 
In this study, parameters for an approximate calculation of the received signal strength were obtained and applied to develop a performance simulation tool for a medium-frequency R-Mode system. 
As a case study, the signal strength from the  Yeongju transmitter was simulated using the proposed parameters. 
\end{abstract}


\section{Introduction}
The global navigation satellite systems (GNSS), including the GPS \cite{Enge11:Global} of the U.S. and Galileo of Europe, are widely used to obtain the locations of users based on the received signals from satellites. 
However, GNSS is susceptible to radio frequency interference \cite{Park2021919, Park2018387, Kim2019, Schmidt20, Park20173888} and ionospheric anomaly \cite{Jiao15:Comparison, Lee17:Monitoring, Seo20111963, Lee22:Optimal, Sun21:Markov, Sun2020889, Ahmed20171792}.
GNSS interference cases have been reported worldwide, and South Korea has been subjected to North Korea’s intentional GPS jamming on multiple occasions since 2010 \cite{Kim22:First, Son20181034}.

South Korea is developing an R-Mode system, a terrestrial integrated navigation system, as one of the backup navigation systems for ships if the GNSS fails \cite{Son22:Analysis, Han2021:R-Mode, Jeong22:Preliminary}. 
There are numerous studies regarding positioning and navigation under GNSS outages \cite{Lee22:SFOL, Park2020800, Lee2020939, Lee20202347, Jeong2020958, Kang2020774, Kang21:Indoor, Park21:Indoor, Lee22:Evaluation, Lee22:Urban, Jia21:Ground, Kim21:GPS, Lee2020:Preliminary, Lee20191187, Kang20191182, Kim2017348}. 
R-Mode utilizes the existing radio signals of opportunity for obtaining position, navigation, and time (PNT) information. 
Compared with eLoran \cite{Son20191828, williams2013:uk, pelgrum2006:new, li2020:research, Kim2020796, Park2020824, Hwang2018}, R-Mode can provide PNT at a relatively lower investment cost for a new radio navigation infrastructure \cite{Johnson2020:R-Mode, Johnson2014:feasibility, Johnson2014:feasibility3, Johnson2014:feasibility1, Johnson2014:feasibility2, Johnson2017:initial}. 
Unlike the eLoran system, R-Mode utilizes a medium-frequency (MF) differential GNSS (DGNSS) signal or a very high frequency (VHF) automatic identification system (AIS) signal \cite{Johnson2014:feasibility1,  Johnson2014:feasibility2, Johnson2014:feasibility3}.

To build an R-Mode system in Korea, a simulation tool that predicts the positioning performance of the system is useful.  
The simulation tool that we are developing will calculate the signal strength and noise value at a given location and predict the positioning performance of the system based on the calculated signal-to-noise ratio (SNR) value. 
Simulation tools for predicting eLoran performance have been developed \cite{Rhee21:Enhanced, Lo08:Loran}; however, a simulation tool for predicting MF or VHF R-Mode performance requires further development \cite{Jeong21:Development}. 
In this study, parameters for an approximate calculation of the received signal strength were obtained and applied to develop a performance simulation tool for an MF R-Mode system.

\section{Methodology}

\subsection{Approximate signal strength calculation}

Lo \textit{et al.} \cite{Lo08:Loran} utilized \eqref{eqn:ss_eLoran} to calculate the eLoran received signal strength at a given location. 
There are other formulas with better accuracy but they require extensive computation time.
Thus, \eqref{eqn:ss_eLoran} is useful approximation for fast computation with reasonable accuracy. 
\begin{equation}
\label{eqn:ss_eLoran}
\begin{split}
    &\mathrm{signal \; strength \; [dB(\mu V/m)]}  \\
    &= 189.353 - 10 \log {r^2} - \sum_{i} {ea}_i r_i
\end{split}
\end{equation}
Here, ${ea}_i$ denotes the extra signal attenuation per unit distance at grid $i$, $r_i$ denotes the signal travel distance in meters within grid $i$, and $r$ denotes the signal travel distance from a transmitter to a receiver in meters.
Thus, ${ea}_i r_i$ is the extra attenuation at grid $i$, and $\sum_{i} {ea}_i r_i$ is the total extra attenuation along the signal propagation path from a transmitter to a receiver. 
This is called \textit{extra attenuation} because it is additional signal attenuation to the $r^2$ attenuation.
This extra attenuation is mainly affected by the ground conductivity along the propagation path. 

The ${ea}_i$ can be obtained based on the results from GRWAVE \cite{Garcia:Calculation}.
GRWAVE calculates signal strength according to the given ground conductivity and center frequency, which was used in the previous eLoran simulator study \cite{Rhee21:Enhanced}. 
Although the results from GRWAVE are reliable, the approximate formula in \eqref{eqn:ss_eLoran} provides significantly faster results. 
Fast computation is important when we need to calculate the signal strengths at numerous grid points over Korea.



\subsection{MF R-Mode signal strength calculation}

\begin{figure}
\centering
   \includegraphics[width=1.0\linewidth]{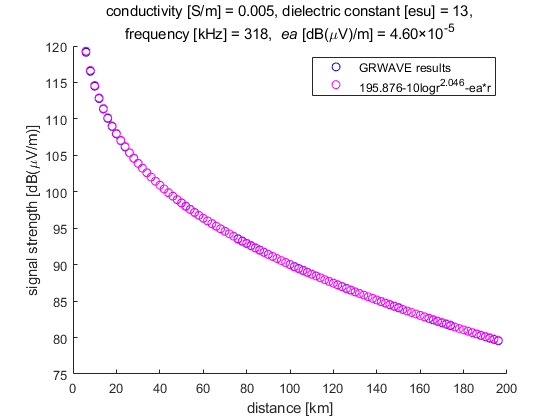}
\caption{Comparison of MF R-Mode signal strength simulation results from GRWAVE and our approximate formula when the ground conductivity is 0.005 S/m.}
\label{fig:GRWAVE}
\end{figure}

We use the same idea with \eqref{eqn:ss_eLoran} of the eLoran case; however, a constant $C$ and an exponent $e$ need to be determined in addition to ${ea}_i$ for MF R-Mode signal strength calculation as shown in \eqref{eqn:ss_R-Mode}. 
\begin{equation}
\label{eqn:ss_R-Mode}
\begin{split}
    &\mathrm{signal \; strength \; [dB(\mu V/m)]}  \\
    &= C - 10 \log {{r}^e} - \sum_{i} {ea}_i r_i
\end{split}
\end{equation}
The $C$ and $e$ are fixed values but ${ea}_i$ depends on the ground conductivity. 
When the ground conductivity is 0.005 S/m, for example, GRWAVE provides the signal strength according to the propagation distance, which is shown as the blue circles in Fig. \ref{fig:GRWAVE}.
The pink circles represent the signal strength from the approximate formula of \eqref{eqn:ss_R-Mode2}, which is the special case of \eqref{eqn:ss_R-Mode} when the ground conductivity is constant over all grids. 
\begin{equation}
\label{eqn:ss_R-Mode2}
\begin{split}
    &\mathrm{signal \; strength \; [dB(\mu V/m)]}  \\
    &= C - 10 \log {{r}^e} - ea \cdot r
\end{split}
\end{equation}
The pink circles in Fig. \ref{fig:GRWAVE} are almost identical to the blue circles when $ea$ is  4.60$\times10^{-5}$ dB($\mu$V/m).
This $ea$ value was obtained by curve fitting in the least squares sense. 

The $C$ and $e$ values of \eqref{eqn:ss_R-Mode} and \eqref{eqn:ss_R-Mode2} were also obtained by curve fitting.
$C = 195.876$ and $e = 2.046$ provide the minimum least squares error over all the ground conductivity cases of Table \ref{tab:ea_cond}.
The $ea$ values that we obtained for various ground conductivity values are summarized in Table \ref{tab:ea_cond}.


\begin{table}
\centering
\caption{Extra attenuation per unit distance according to the ground conductivity when $C$=195.876 and $e$=2.046}
\label{tab:ea_cond}
\begin{tabular}{c|ccc}
\noalign{\smallskip}\noalign{\smallskip}\hline\hline
conductivity
[S/m] 
& $ea$ for MF R-Mode signal 
[dB($\mu$V)/m]
 \\
\hline
5$\times10^{-4}$ & 2.24$\times10^{-4}$
 \\
\hline
1$\times10^{-3}$ & 1.64$\times10^{-4}$
 \\
\hline
2$\times10^{-3}$ & 1.04$\times10^{-4}$
 \\
\hline
5$\times10^{-3}$ & 4.60$\times10^{-5}$
 \\
\hline
8$\times10^{-3}$ & 2.89$\times10^{-5}$
 \\
\hline
1$\times10^{-2}$ & 2.37$\times10^{-5}$
 \\
\hline
4 & -5.40$\times10^{-7}$
 \\
\hline
\end{tabular}
\end{table}


\section{Results and Discussions}

\begin{figure}
\centering
   \includegraphics[width=1.0\linewidth]{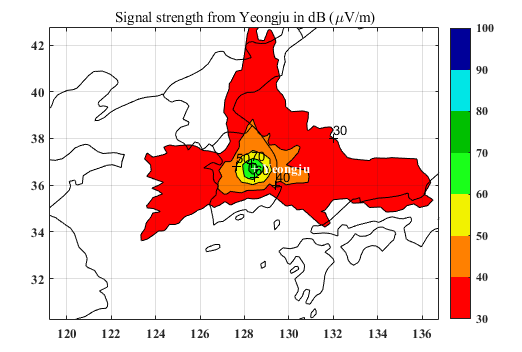}
\caption{MF R-Mode signal strength simulation result for the Yeongju transmitter.} 
\label{fig:Yeongju}
\end{figure}

The result of MF R-Mode signal strength simulation over Korea with the parameters of Table \ref{tab:ea_cond} is presented in Fig. \ref{fig:Yeongju} when the transmitter is in Yeongju. 
The actual coordinates of the current Korean MF R-Mode testbed transmitter in Yeongju were used. 
As proposed in \cite{Rhee21:Enhanced}, a land cover map was used to determine the ground conductivity of the Korean region.



Considering the different $ea$ values for different ground conductivity listed in Table \ref{tab:ea_cond}, the signal strength of the MF R-Mode system significantly varies depending on the  conductivity of the area.
Thus, the signal strength in Fig. \ref{fig:Yeongju} does not uniformly decrease from the Yeongju transmitter.

\section{Conclusion}

In this study, the parameters (i.e., $C$, $e$, and $ea$) of the approximate signal strength calculation formula for the MF R-Mode system were determined and applied to the simulation tool. 
The parameters that minimize the least squares error between the results from the GRWAVE and approximate formula were selected. 
As a case study, the signal strength of the MF R-Mode transmitter in Yeongju was simulated over Korea using the proposed parameters.

\section*{Acknowledgment}

This research was conducted as a part of the project titled ``Development of integrated R-Mode navigation system [PMS4440]'' funded by the Ministry of Oceans and Fisheries, Republic of Korea (20200450).

\bibliographystyle{IEEEtran}
\bibliography{mybibfile, IUS_publications}

\vspace{12pt}

\end{document}